\documentclass[12pt]{article}
\usepackage{amsmath}
\usepackage{hyperref}
\usepackage{graphicx}

\usepackage{enumerate}
\usepackage{url} 
\usepackage[utf8]{inputenc}
\usepackage{amsmath, amsfonts, mathrsfs}
\usepackage{enumitem}
\usepackage{wrapfig}
\usepackage{natbib}

\usepackage{indentfirst}
\usepackage{placeins}
\usepackage{xcolor}
\usepackage{comment}
\usepackage{float}
\usepackage{authblk}
\usepackage{caption}
\usepackage{microtype}

\newcommand{\blind}{0}

\begin{document}

\def\spacingset#1{\renewcommand{\baselinestretch}%
{#1}\small\normalsize} \spacingset{1}


\if0\blind
{
 \title{\bf Mathematical Analysis of Redistricting in Utah}

\author[1]{Annika King}
\author[1]{Jacob Murri} \author[1]{Jake Callahan}
\author[1]{Adrienne Russell}
\author[1]{Tyler J. Jarvis}

\affil[1]{Department of Mathematics, Brigham Young University}

\date{July 13, 2022}

 \maketitle
} \fi

\if1\blind
{
 \bigskip
 \bigskip
 \bigskip
 \begin{center}
 {\LARGE\bf Title}
\end{center}
 \medskip
} \fi

\bigskip
\begin{abstract}
We discuss difficulties of evaluating partisan gerrymandering in the congressional districts in Utah and the failure of many common metrics in Utah.
We explain why the Republican vote share in the least-Republican district (LRVS) is a good indicator of the advantage or disadvantage each party has in the Utah congressional districts. 
Although the LRVS only makes sense in settings with at most one competitive district, in that setting it directly captures the extent to which a given redistricting plan gives advantage or disadvantage to the Republican and Democratic parties. We use the LRVS to evaluate the most common measures of partisan gerrymandering in the context of Utah's 2011 congressional districts. We do this by generating large ensembles of alternative redistricting plans using Markov chain Monte Carlo methods.  We also discuss the implications of this new metric and our results on the question of whether the 2011 Utah congressional plan was gerrymandered. 
\end{abstract}

\noindent%
{\it Keywords:} Markov chain Monte Carlo, redistricting, Utah, ensemble methods, gerrymander, partisan symmetry
\vfill

\newpage
\spacingset{1.45} 
\section{Introduction}
\label{sec:intro}

It has been claimed that the U.S.~congressional districts in Utah that were enacted in October 2011  constituted an unfair Republican gerrymander \citep{trib, Magleby, A4BU}. 
Analysis of these claims is complicated by the fact that some of the commonly used measures of partisan gerrymandering suffer from what is called the \emph{Utah paradox}: these metrics flag redistricting plans in Utah with more Democratic representation as egregious pro-Republican gerrymanders and flag plans with no Democratic representation as pro-Democratic gerrymanders \citep{DDDGMSS}.

In this paper we propose a new measure of partisan gerrymandering and examine it and the other most common measures of gerrymandering 
in the context of Utah's 2011 congressional districts. We do this by generating large ensembles of alternative redistricting plans using Markov chain Monte Carlo methods, which have become standard in the study of redistricting. We use data from the 2010 general election in Utah because that best represents the partisan distribution of voters in Utah at the time the enacted redistricting plan was adopted in 2011. 

In order to understand different metrics applied to Utah, we need to understand characteristics of Utah and its enacted congressional districts that complicate the analysis. 
Most of the Democratic voters in Utah reside in the greater Salt Lake area, which was divided among three districts. Each district had a Republican majority, despite the fact that Utah had an approximately $65/35$ Republican-Democrat split and a Democratic congressman at the time the plan was enacted. This might suggest the plan was designed to unfairly favor the Republican party.

\begin{figure}
 \centering
\includegraphics[width=0.4\textwidth]{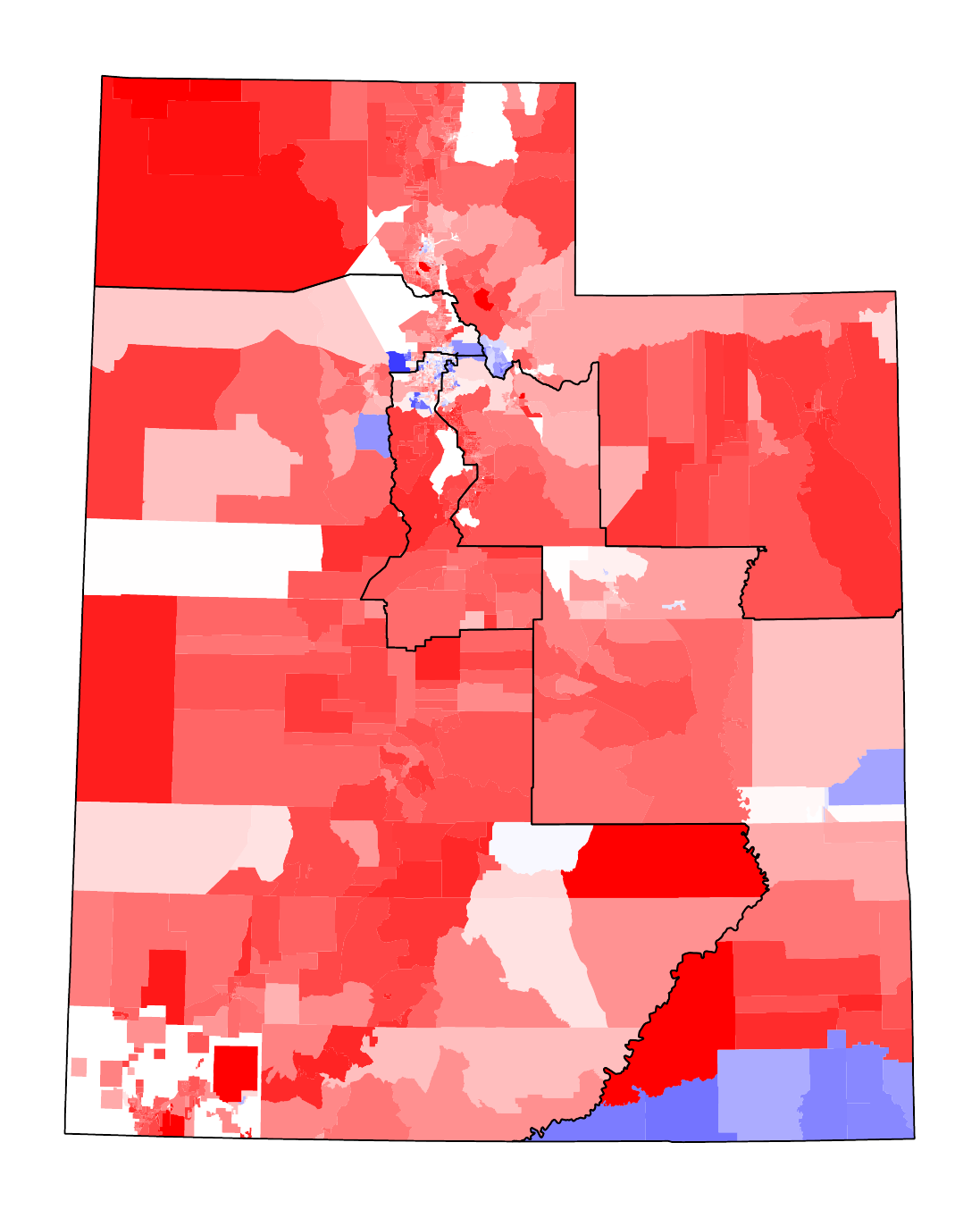}
 \includegraphics[width=0.4\textwidth]{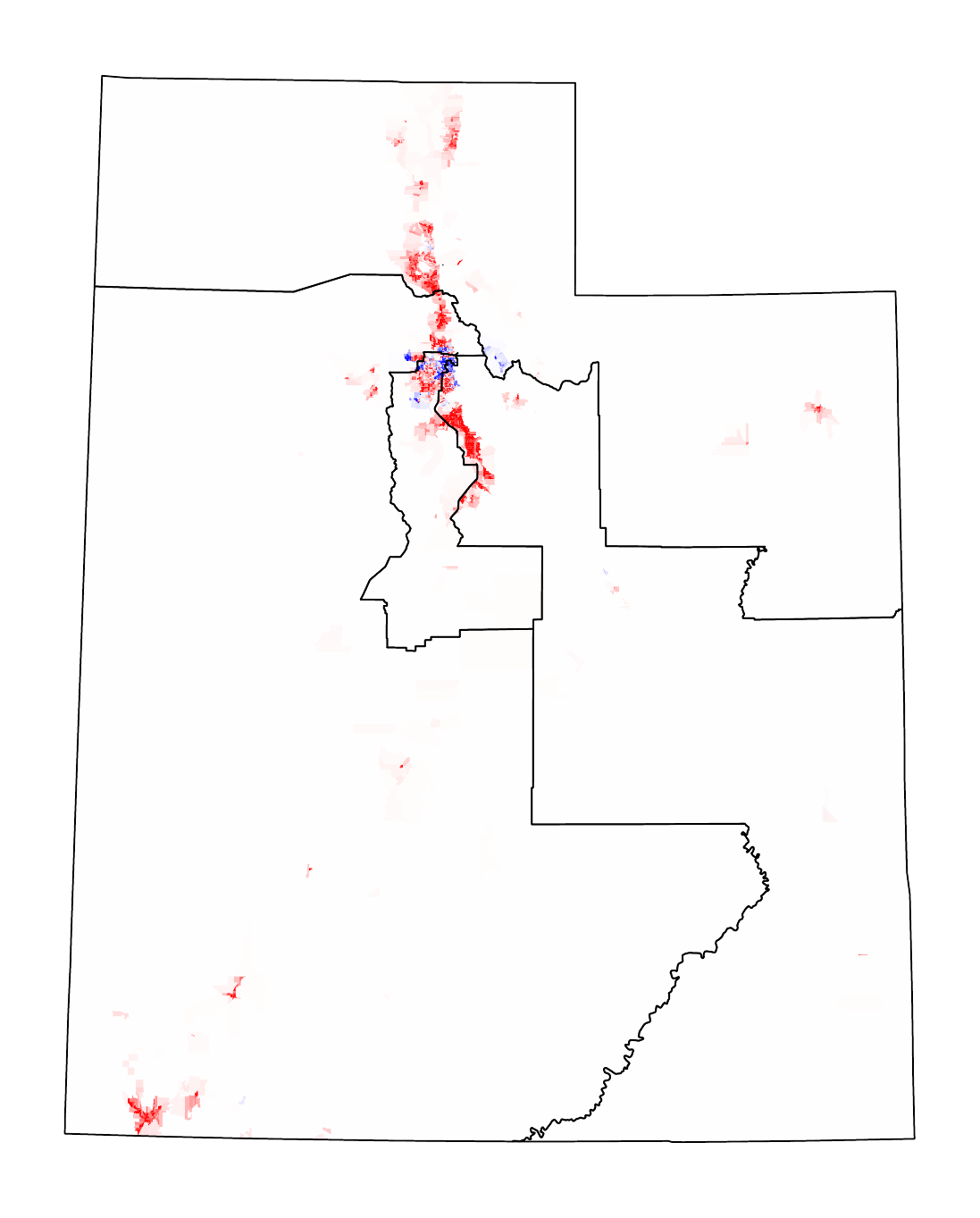}
 \caption{Two plots of the four Utah congressional districts, overlayed on 
 representations of the political geography of the state. The typical plot (left panel) of Utah political geography, with each precinct colored by the percentage of Republican or Democratic voters, is misleading, because many of Utah's precincts are very large and mostly uninhabited. The plot in the right panel is more accurate and informative; here the color indicates vote difference (number of Republican votes minus the number of Democratic votes) \emph{divided by the area of the precinct}. This plot makes it easier to see how the partisan vote share is distributed across the state. The fact that most of the population resides in the narrow corridor of the Wasatch Front is also clearly visible.}
 \label{fig:VoteByArea}
\end{figure}

But more than one third of Utah’s population lives in Salt Lake County, and all four congressional districts must be nearly equal in population, meaning that Salt Lake County must be split between at least two congressional districts. Additionally, roughly 80\% of Utah's population lives in the relatively small area known as the \emph{Wasatch Front} in the north-central part of the state, not far from Salt Lake City. The rest of the state is sparsely populated, with large empty regions, and the remainder of the population is mostly concentrated in a few municipalities like St.~George, Logan, and Cedar City (see Figure~\ref{fig:VoteByArea}). This means that each of the four congressional districts must contain part of the Wasatch Front in order to maintain population equality, and thus it might not be unreasonable for each of the four districts to take a piece of Salt Lake County.

Many metrics for quantifying gerrymandering depend, at least in part, on the percentage of seats expected to go to each party.
But the small number of competitive congressional seats in Utah means that the number of Republican or Democratic seats in a given plan is not a meaningful indicator of how gerrymandered the plan. Indeed, using the 2011 voting precincts and 2010 election returns, it does not appear to be possible to draw a redistricting map of Utah that has two districts with a majority Democratic vote share (see Section~\ref{sec:unusual} for details). This means redistricting plans in Utah at that time could only have zero or one Democratic seat, which gives reason to be suspicious of metrics that rely on the number of seats. 

The arguments above suggest that evaluating a redistricting plan in Utah has some subtle and interesting aspects, and that many popular metrics used in the study of gerrymandering might be misleading when used in Utah.

\subsection{Previous Work}\label{sec:previous}

\subsubsection{Ensembles for Evaluating Redistricting Plans}\label{sec:ensemble-background}

Many different metrics for evaluating redistricting plans have been proposed, and in many cases the proposers also gave an approximate range of values that would indicate a gerrymander. But different states have differing political geographies, and the political geography strongly influences what the possible range of values for a given metric might be. In order to use any metric to evaluate a redistricting plan, the values taken by the metric must be considered in the context of what is actually possible in the given setting. To do this it is necessary to create a large ensemble of alternative plans, constructed without consideration of political data, and drawn in a way that is representative of the distribution of possible or acceptable plans \citep{MattinglyNC, DDDGMSS, Pegden, Cho, ChenRodden, MaglebyMosesson}. If the enacted plan is an outlier, that could mean either that the enacted plan is a political gerrymander or that the policies that were used to produce the plans in the ensemble differ significantly from those used to draw the enacted plan.

The two main methods for constructing alternative plans are \emph{agglomerative} methods and \emph{Markov chain Monte Carlo (MCMC)} methods. Agglomerative methods build districts by incrementally joining precincts into districts. 
Examples of these methods include \citet{ChenRodden}, \citet{Cho}, \citet{Magleby}, and \citet{MaglebyMosesson}.

Agglomerative methods have several disadvantages. First, they are relatively slow, meaning they usually cannot produce enough plans to sample the space of possible plans very well. Moreover, these methods often produce the same plan repeatedly. More importantly, it is not known which distribution the agglomerative methods sample from, and there is some evidence that they tend to sample in a way that many would call biased \citep{Fifield,BeckerSolomon}. 

MCMC methods for analysis of redistricting begin with a given starting plan and use that to construct a new plan in some way---usually either by ``flipping'' a precinct out of one district and into another adjacent district, or by merging two districts into one large district and then redividing the merged district in a different way (dividing a merged district into two districts of equal population is much easier, computationally, than dividing it into three or more districts). Some important examples of the use of MCMC for this sort of analysis include \citet{MattinglyNC}, \citet{MattinglyWisc}, \citet{DDDGMSS}, \citet{DDSRecombination}, \citet{Pegden}, and \citet{Fifield}.

MCMC methods must begin with an existing plan but they are fast and can be designed to sample from specific distributions. They also generally produce few duplicates. Moreover, when MCMC methods run long enough, duplicate plans, while not necessarily very common, are informative, indicating plans that are more probable in the stationary distribution. Because of the many advantages of MCMC methods over agglomerative methods, we use MCMC methods in this paper.

\subsubsection{Previous Work on Gerrymandering Metrics in Utah}\label{sec:previous_utah}

We are aware of two other studies of  gerrymandering metrics in the context of the 2011 U.S. Congressional districts in Utah, both using election returns from the 2016 general election. The first is an unpublished report by \citet{Magleby}, commissioned by the Fair Redistricting Caucus of Utah, to evaluate whether Utah's Congressional districts had been gerrymandered. The second is the paper of \citet{DDDGMSS}, which was focused primarily on paradoxical properties of gerrymandering metrics when applied to some states, including Utah.

Magleby uses an agglomerative algorithm that he and Mosesson (\citeyear{MaglebyMosesson}) developed to build an ensemble of 10,000 redistricting plans in Utah. 
Arguing that the division of heavily Democratic Salt Lake County into three districts appears to be a cracking gerrymander, 
distinguished by a relatively uniform distribution of Democrats across all districts, he uses the standard deviation of the Democratic vote share in the four districts as his measure of the severity of the gerrymander, with a smaller standard deviation indicating a more severe cracking gerrymander. For each of the different races he found that the standard deviation for the enacted plan was lower than it was for any of the other plans in his ensemble, and concluded that the enacted plan was an ``egregious gerrymander.''

\citet{DDDGMSS} used MCMC methods (the \emph{ReCom} method implemented in the \emph{GerryChain} package \citep{gerrychain}) to generate 100,000 plans for Utah and analyzed several gerrymandering metrics using data from the 2016 senate race (Lee v.~Snow). They observe first that under the vote distribution of the 2016 senate race, it appears to be impossible to make a plan with more than one seat going to the Democrats, and that most ($94\%$) of the plans generated have all four seats going to the Republicans. They also observe what they call the \emph{Utah paradox}, where common partisan symmetry scores (partisan Gini, partisan bias, and mean--median score) are supposed to indicate fairness when the score is close to zero and unfairness when the score is farther from zero, and yet these scores were close to zero in their ensemble only when all four seats go to Republicans. Moreover, the partisan bias and mean--median scores make a sign error, meaning that they indicate a pro-Republican gerrymander on all plans that give a seat to the Democrats.

\section{Methods and Data}
\subsection{A New Metric to Measure Gerrymandering in Utah}\label{sec:unusual}

Measuring gerrymandering and partisan fairness in Utah requires extra care, as we now describe. 

\subsubsection{At Most One Democratic Seat}\label{sec:at-most-one}

Using 2011 precincts and 2010 election returns, it appears to be impossible to construct a valid redistricting plan in Utah with two or more majority-Democratic districts. We have randomly generated over 200 million redistricting plans with contiguous districts, and not one of them gives two seats to the Democrats. With some computational work (see the Appendix), using the partisan distribution of the 2010 senate election returns, we were able to construct a redistricting plan that gives two of the four congressional seats to the Democrats and still complies with the population bound (less than 1\% deviation from population equality), but we were only able to do this by removing the requirement that the districts be contiguous.  
It appears extremely unlikely that this could be accomplished under the additional requirement of contiguity. 

The immediate practical result of this is that in Utah one cannot get a very useful measure of partisan gerrymandering just by counting the number of districts with a Republican or Democratic majority. This suggests that many popular measures of partisan symmetry will not give good information in Utah.

\subsubsection{Least Republican Vote Share (LRVS)}\label{sec:LRVS}

In any given plan the vote share in the least-Republican district determines whether Utah elects three or four Republicans to Congress. But even when the Republican vote share in this least-Republican district is greater than $50\%$, as it was in the 2011-enacted plan, that does not guarantee the Republicans will win the seat. Although the least-Republican district in Utah has consistently voted majority Republican in senate and gubernatorial elections for the past ten years, it has elected a Democrat to the House of Representatives in two of the past five elections: Matheson in 2012 and McAdams in 2018. 
 
Plans with a greater share of Republican voters in the least-Republican district have a greater probability of electing a Republican in that district (that is, of electing four Republicans to Congress instead of three). Since Democrats cannot win two seats, the most they could hope for is a better chance at winning the seat in the least-Republican district. This suggests the use of the Republican vote share in the least-Republican district as an indicator of how pro-Republican a redistricting plan is in Utah. We call this measure the \emph{Least-Republican Vote Share (LRVS)}. LRVS is a number between $0$ and $1$, corresponding to percentage of Republican voters in the least-Republican district.  The higher the LRVS, the more likely it is that the one competitive seat will go to a Republican, and the lower it is, the more likely that seat will go to a Democrat.  

The LRVS would not be meaningful in most other states or even in other types of districts in Utah; Section~\ref{other-states} discusses other states where it could be useful.  
But for the Congressional districts in Utah, the LRVS gives a very clear indication of how pro-Republican or pro-Democratic a given plan is, while, unfortunately, most other metrics do not. As shown by \citet{DDDGMSS} using 2016 senate returns, some common metrics of gerrymandering fail spectacularly when applied to Utah, mislabeling plans that give more seats to Democrats as  pro-Republican gerrymanders. \citet{Veomett} has shown that another common metric, the efficiency gap, is deeply problematic in many settings. Some problems with several other common metrics have been identified by \citet{Warrington}. One of the main questions we seek to answer in this paper is which metrics other than the LRVS give correct and useful information about possible gerrymandering of the U.S.~congressional districts in Utah. 

There is no prescribed or ideal value for the LRVS.  Reasonable values of the LRVS can only be  determined in comparison to  the marginal distribution of LRVS values for acceptable plans, which we approximate by constructing large ensembles drawn from appropriate distributions.  Of course, the range of acceptable values for most, if not all, metrics should be determined in this way, despite any sort of hypothetical partisan symmetry that specific values may be thought to represent.

\subsection{Data}
\label{sec:data-and-methods}
\label{sec:data}

We evaluated the U.S.~Congressional districts in Utah, enacted in October 2011, using election returns from the 2010 senate and gubernatorial elections. 

\subsubsection{Justification for Using 2010 Returns}

Both of the two previous papers analyzing the Utah congressional districts \citep{DDDGMSS, Magleby} use returns from the 2016 election. But the demographics in Utah changed rapidly from 2011 to 2016, and the returns from the 2016 election, even if they accurately reflected voter preferences in 2016, may not accurately reflect the partisan composition of the state at the time the districts were drawn. The 2010 election returns provide the latest data set the state legislators had available at the time they were creating these districts. 

The demographic shift alone would justify preferring the 2010 returns instead of the 2016 returns, but there are several other reasons why the 2016 returns are problematic. First, in 2016 every statewide race, other than the presidential race, involved an incumbent Republican running for reelection. The well-documented incumbent advantage \citep{incumbent89,GelmanKing,incumbent06} in American politics suggests the returns in these elections are probably not representative of the electorate's underlying preferences for Democrats or Republicans. 

Second, the senate race that year, which was used by \citet{DDDGMSS} in their analysis, was especially unusual, since the Republican incumbent, Mike Lee, was challenged by Misty Snow (D), one of the first two openly transgender candidates nominated by a major political party for a federal office in America. It is likely, in the socially conservative setting of Utah, that this resulted in many more votes for  Lee than we would otherwise expect from the incumbent advantage alone. 
Third, the state auditor race was also unusual, in that the Democratic candidate dropped out, leaving the incumbent Republican essentially unopposed. 

Finally, the presidential election in 2016 was unusual because the Republican candidate, Donald Trump, was so unpopular in Utah that an independent candidate, Evan McMullin, received a significant share of the vote ($21.5\%$), much of which appears to have come from otherwise-Republican voters. Compare this to other elections in Utah where independents and other party candidates typically get no more than $2\%$--$5\%$ of the vote share.

\subsubsection{Statewide Races in 2010}

Utah had two statewide races in the 2010 general election, namely a senate race and a gubernatorial special election. The senate race of Mike Lee (R) v.~Sam Granato (D) had no incumbent in the race and at the time neither candidate stood out as especially unusual for his party. In that race 390,179 votes were cast for Lee and 207,685 for Granato (approximately a $65/35$ split, ignoring other parties). A Constitution Party candidate also received 35,937 votes ($5\%$ of the vote), but that party does not seem to have had any practical effect on the outcome of races in Utah.

The gubernatorial race was a special election to replace the popular Jon Huntsman (R), who stepped down to serve as ambassador to China under a Democratic president, Barack Obama. His lieutenant governor, Gary Herbert (R), who had been acting governor for several months already, could almost be considered an incumbent in this race. Herbert won the election with 412,151 votes to his Democratic opponent Corroon's 205,246 votes (about a $67/33$ split, ignoring other parties). An Independent and a Libertarian candidate each picked up about $2\%$ of the vote, but again, these parties have not seemed to have a practical effect on the outcome of races in Utah (with the notable exception of the 2016 presidential election), so we focus our analysis on just the two main parties, Republicans and Democrats.

The party support was distributed differently across the state in these two statewide elections. For example, in Congressional District 2, there was a $61/39$ split using the senate returns and a $70/30$ split using the gubernatorial returns. The discrepancies within and across the districts result in differing conclusions about partisan advantage. 

We analyze and report our results using both the senate and gubernatorial returns, and we also consider the results of averaging the partisan distributions of the two races.  But we believe the senate returns give the most accurate representation of voting preferences in 2010 because the gubernatorial election was a special election, and the Republican, Herbert, was, in many ways,  effectively an incumbent.

\subsubsection{Data Sources and Preparation}

District and precinct boundaries, as well as 2010 demographic information from the U.S. Census, were retrieved from the official Utah State Automated Geographic Reference Center \citep{UtahAGRC:districts,UtahAGRC:precincts, UtahAGRC:census}. Precinct-level election returns for the 2010 general election in Utah were provided to us by the Utah Lieutenant Governor's office.

The data required significant cleaning to make them usable for analysis, including making election returns machine readable and uniformizing their format across counties. Precinct geometries required some correction to remove large gaps and overlaps. Some precincts consisted of multiple disconnected pieces. These disconnected precincts were merged with neighboring precincts to make the combined precinct connected. A few precincts were also completely contained inside another precinct. These precincts were merged with the precinct that contained them. Merging precincts in these ways amounts to requiring that certain small groups of precincts can only be assigned to a district as a group. The space of redistricting plans where these groupings are preserved is a strict subset of the space of all possible redistricting plans, but merging is necessary to ensure that every plan in the sample space is contiguous. Merging reduced the total number of precincts by 331, from 2,974 to 2,643.

\subsection{Creating Ensembles}
\label{sec:MCMC-methods}

\begin{figure}[h]
 \centering
 \includegraphics[width=0.85\textwidth]{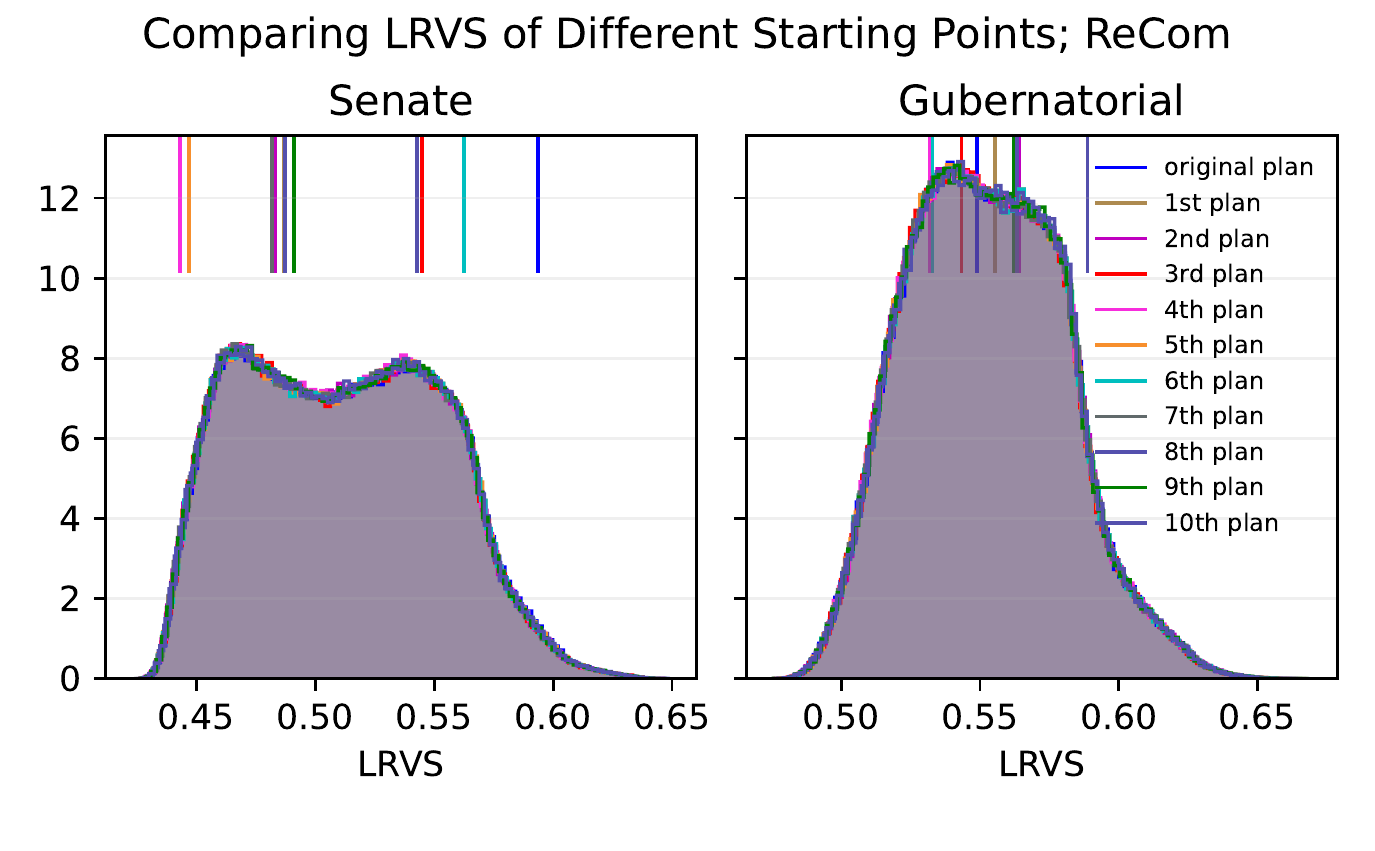}

 \caption{
 Density plots for the LRVS using ReCom chains starting at eleven different plans (the enacted plan and ten others). Each chain was run for one million steps.
 The vertical lines represent the LRVS of the starting points. In the left column the LRVS is computed using the 2010 senate race returns, and in the right column the densities are computed using the 2010 gubernatorial race returns. In both cases all eleven densities appear to be very similar, suggesting that the chains have mixed well, at least in terms of their effect on the LRVS.}
 \label{fig:hist_recom}
\end{figure}

We use Markov chain Monte Carlo (MCMC) methods to generate large ensembles of alternative redistricting plans. We initially tried three different MCMC methods, and our code implementation for all three of these methods was based on the open source \emph{GerryChain} code base \citep{gerrychain} in Python, with only minor adjustments, including adding a hash function to identify how many of the plans generated by the various methods were duplicates.
The three MCMC methods we used were \emph{uniform flip}, \emph{weighted flip}, and \emph{ReCom}. Some basic chain-mixing analysis showed that the \emph{ReCom} chains mixed much better than the two flip-based methods. In particular, the Gelman--Rubin potential scale reduction factor (PSRF) test was done for each chain. A PSRF between $1.00$ and $1.01$ is generally taken to indicate good convergence; whereas a PSRF larger than $1.01$ indicates that the chains are not mixing well. All of the ReCom chains had PSRF between $1.00$ and $1.01$, but none of the flip chains did. Moreover, density plots for each of the chain types beginning at different starting points showed that the flip-based chains did not mix well, but the ReCom chains did (see Figure~\ref{fig:hist_recom}). For these reasons we use ReCom in this paper rather than uniform and weighted flips. 

The flip-based methods have known stationary distributions, whereas the stationary distribution for ReCom is unknown, but there is some evidence that the ReCom stationary distribution is very similar to the spanning tree distribution \citep{DDSRecombination}.  Moreover, ReCom is a widely used method that tends to create relatively compact districts.

An important question about the choice of an ensemble is whether the policies used to construct the ensemble are similar to the policies used, or supposed to be used, to draw the enacted plan.  In the case of Utah's 2011 enacted plan, the Legislative Redistricting Committee's adopted redistricting principles \citep{UTpolicy} included only that districts must be contiguous, have nearly equal population, and be ``reasonably  compact.''  
All the MCMC methods we used draw plans that are contiguous and have nearly equal population, so the primary question for us is how to match the Legislative Committee's idea of ``reasonably compact.''  

The concept of reasonably compact was not defined by the Committee, but it is generally agreed that most plans drawn by uniform flip methods are not compact.  Weighted flip methods require choices of what measure to use for compactness and how strongly the distribution should prefer plans that score well by that measure (via a Gibbs energy function). Stronger preference for good compactness scores tends to lead to slower convergence to the stationary distribution.  In our experiments, none of our choices of compactness measures and weights resulted in good convergence using weighted flips, probably because the proposal distribution we used (uniform flip) has an overwhelming preference for noncompact districts (see \citet{DDSRecombination}). 

With all types of our chains we considered using an isoperimetric bound like Polsby--Popper, but those scores penalize districts with meandering or snaky boundaries, such as often occur along natural boundaries like the Green and Colorado river.   That means that districts that respect natural boundaries are less likely to be chosen under a distribution that bounds isoperimetric scores.  Discouraging splits along natural boundaries was not a stated policy preference of the Redistricting Committee in 2011, and in 2021 the adopted statute guiding redistricting stated a preference to encourage splits along natural boundaries.  This suggests that an isoperimetric bound would be inappropriate in our analysis. 

While our chains using flip-based methods did not produce compact districts, ReCom and the spanning tree distribution naturally favor districts that many people agree are reasonably compact \citep{DDSRecombination}, and they do so without any user-defined inputs or constraints.  The ensemble we use in this paper was constructed using ReCom with no constraints other than approximate population equality (plus or minus $1\%$) and contiguity.  For comparison  we also constructed a separate ReCom ensemble with an additional absolute constraint on the number of cut edges---a popular measure of compactness. This constrained ReCom ensemble also mixed well, and the results for both the unconstrained ReCom ensemble and the constrained ensemble were very similar, in terms of LRVS and the other metrics of partisan advantage discussed in this paper.

It is, nevertheless, possible that bounding the number of cut edges could potentially have some policy-level implications that were not part of the Legislative Committee's stated principles.  For example, limiting the number of cut edges might discourage plans that cut along the Wasatch front, and encourage splitting in more rural areas (this example was pointed out to us by a referee of this article).  If that is the case, limiting cut edges could have implications for urban versus rural representation, which was a hotly debated policy choice in the 2021 redistricting process in Utah. 

But the ReCom chains both with and without cut-edge constraints had similar results in terms of the LRVS and other political metrics used in this paper.  That suggests that constraining the number of cut edges does not have a significant impact on the partisan outcomes of redistricting plans in Utah.

\begin{figure}[h]
 \centering
 \includegraphics[width=0.85\textwidth]{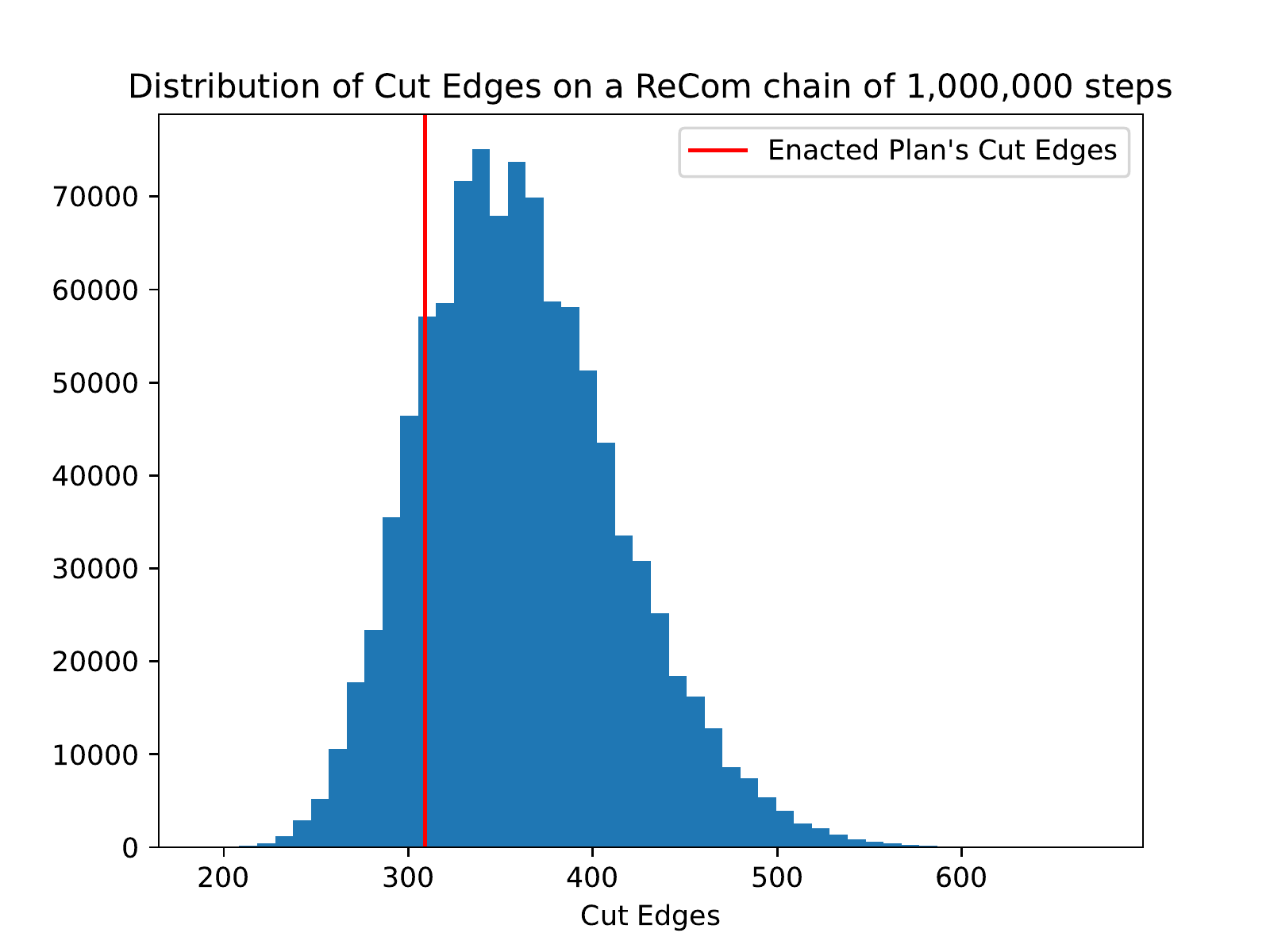}

 \caption{
 Distribution of the number of cut edges in an unconstrained ReCom ensemble of one million plans with no compactness restrictions.  The number of cut edges in the enacted plan (indicated in red) lies within the main body of this distribution, but with somewhat fewer cut edges than the median of the ensemble.  This suggests that the ReCom preference for fewer cut edges is compatible with the Legislative Committee's preference for ``relative compact'' districts.}
 \label{fig:cut_edges}
\end{figure}

Finally, it is possible that the policies used to construct the enacted plan, specifically the Legislative Committee's idea of compactness, might differ substantially from the compactness preferences of the unconstrained ReCom ensemble.  \citet{Clelland} show that ReCom ensembles do prioritize a low number of cut edges, suggesting the possibility that the Legislative Committee might have permitted more cut edges than ReCom.  
This possibility cannot be completely ruled out, but, as shown in Figure~\ref{fig:cut_edges}, the number of cut edges in the enacted plan lies solidly within the main body of the distribution of cut edges in a large unconstrained ReCom ensemble.  This suggests that the unconstrained ReCom preference for fewer cut edges should be  compatible with the Legislative Committee's policy requiring ``relative compact'' districts.

The remainder of the analysis in this paper is based on a ReCom chain of one million steps, with no constraints other than contiguity and approximate population equality, starting at the enacted plan. Using a hash function, we can guarantee that there are no more than $0.0006\%$ duplicate plans in this ensemble.

 At each step (each alternative plan) in the Markov chain, we recorded the partisan composition of each district, as measured by returns from the 2010 senate and gubernatorial races, as well as the values of LRVS and many popular gerrymandering metrics for each plan, including mean--median, partisan bias, partisan Gini, efficiency gap, partisan dislocation, buffered declination, and several other metrics. 
 \FloatBarrier
\section{Results}

\subsection{Least-Republican Vote Share}\label{sec:LRVS_results}

Using the senate data, the enacted plan has an LRVS of $0.594$, which is more Republican than 98.23\% of plans in the ReCom ensemble, suggesting that the enacted plan is a Republican-favoring outlier. But using the gubernatorial data, the enacted plan has an LRVS of $0.5488$, which is close to the median of the gubernatorial LRVS in the ReCom ensemble. Indeed, it is more Republican than only 46.97\% of plans in the ReCom ensemble. 

\begin{figure}
 \centering
	 \includegraphics[width=0.95\textwidth]{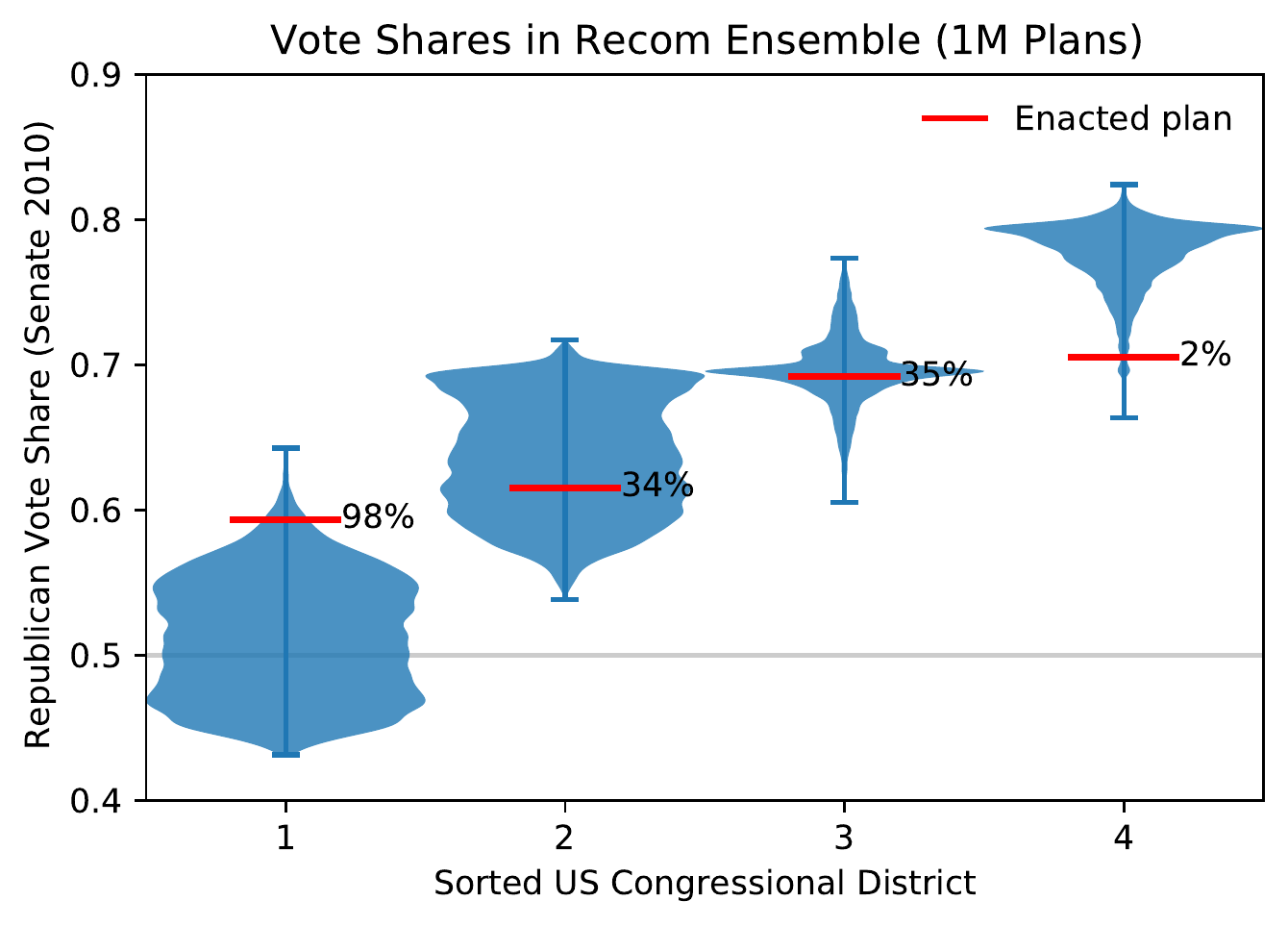}
	 \label{fig:vote_shares_violin}
	 \caption{Distribution of Republican vote shares in districts sorted by vote share. For each plan in the ensemble, the four districts in the plan are sorted from least to greatest Republican vote share (using the senate data). The first (leftmost) violin plot corresponds to the distribution of Republican vote share in the least-Republican district in each plan, that is, the Least Republican Vote Share (LRVS). The red line labeled $98\%$ on that violin indicates that the enacted plan has an LRVS that is higher than $98\%$ of all plans in the ensemble.
	 The fourth (rightmost) violin plot corresponds to the distribution of vote shares in the most-Republican district in each plan, and the line labeled $2\%$ on that violin indicates that only $2\%$ of all plans have a lower Republican vote share in the most-Republican district. }
\end{figure}

\subsection{Measures of Partisan Symmetry}\label{sec:metrics}

We used the unconstrained ReCom ensemble of one million plans to compare the results of the most common measures of gerrymandering and partisan fairness with the LRVS in Utah. The results for each of these metrics are presented below. Our convention is always that the scores are constructed for the Republican party, so that for signed metrics larger positive scores indicate a plan that is more favorable to Republicans.

\subsubsection{Mean-Median Score}

The mean--median score for a given plan is the difference between a party's mean vote share and its median vote share across all the districts in a plan. A large positive mean--median score is supposed to indicate that the plan favors Republicans, and the larger the score, the more the plan supposedly benefits Republicans. 

\begin{figure}
 \centering
 \includegraphics[height=0.8\textheight]{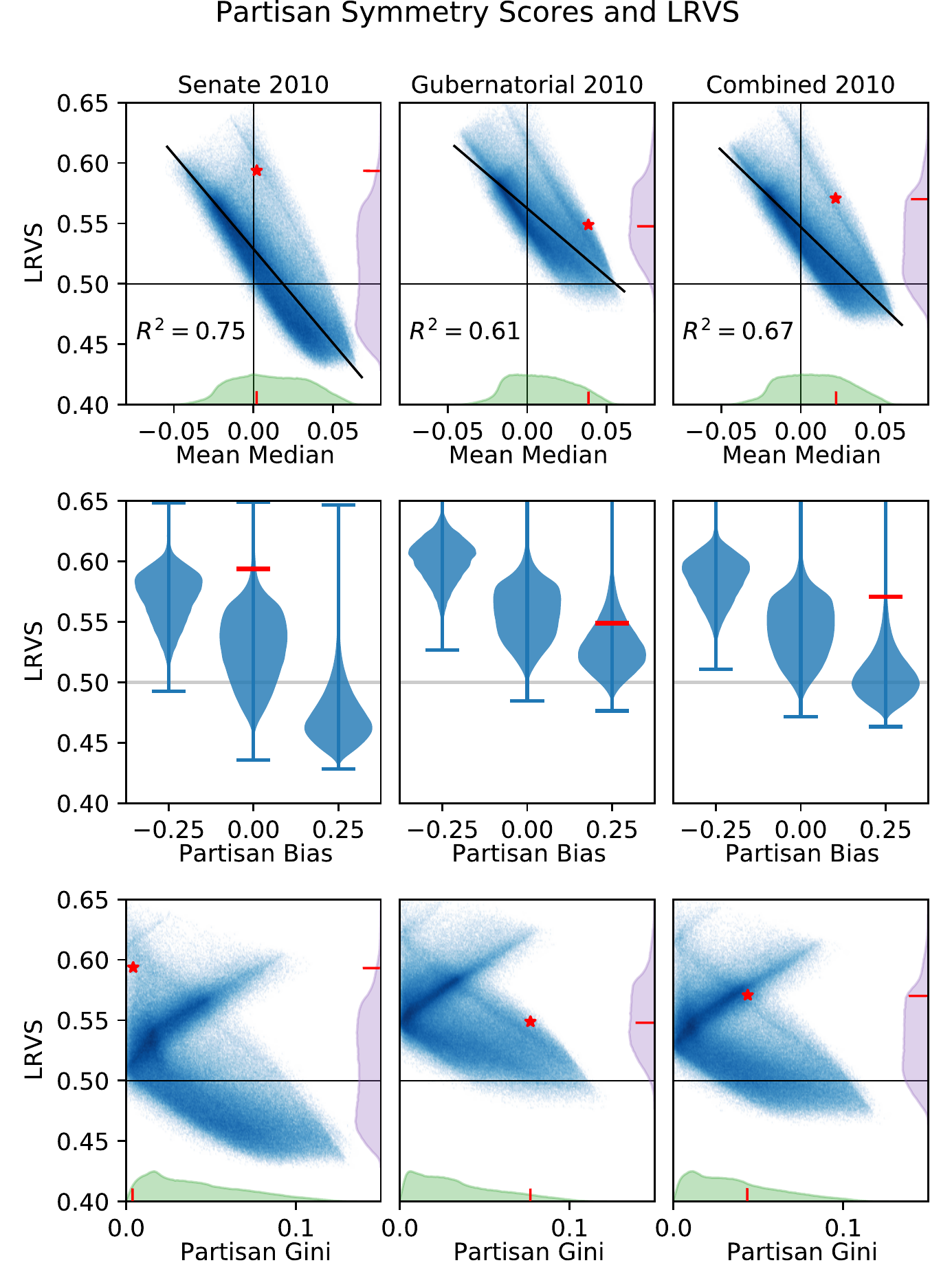}
 \caption{Plots comparing the relationship between three partisan symmetry scores and LRVS in an ensemble of one million plans. The columns correspond to the election returns used to compute the partisan distribution. In each plot the enacted plan is indicated with a red star, and the short red lines in the marginal distributions along the axes indicate where the enacted plan falls in the  marginal distribution.
 The top row shows each of the plans in the ensemble, plotted in blue with its mean--median score (x-axis) and LRVS (y-axis). 
 The middle row shows violin plots of the partisan bias score. 
 The bottom row shows partisan Gini and LRVS. 
 For the mean--median (top row) and partisan Gini (bottom row), the marginal densities for the corresponding metric (green) and for LRVS (purple) are shown on the x and y axes respectively. 
 }
\label{fig:ps_lrvs_recom}
\end{figure}

However, the mean--median score makes what \citet{DDDGMSS} call a \emph{sign error}. The scatter plots in the top row of Figure~\ref{fig:ps_lrvs_recom} show that the the mean--median score is negatively correlated with Republican advantage, rather than the expected positive correlation. 
A linear regression shows a good linear fit for all three data sets (senate $R^2 = 0.75$, Gubernatorial $R^2= 0.61$, and combined $R^2=0.67$), and all three have a negative slope, indicating that a higher LRVS is incorrectly flagged by the mean--median score as being more favorable to Democrats.

\subsubsection{Partisan Bias} 
	
The partisan bias score tries to quantify the gap between a party's (in our case, the Republican party's) seat share and vote share. This makes it a very intuitive measure of gerrymandering. However, in Utah it can only take one of three possible values ($-\frac{1}{4}$, $0$, and $\frac{1}{4}$), so it is relatively uninformative. Like the mean--median score, positive scores are meant to indicate a more Republican-favoring plan. But, as with the mean--median score, partisan bias is negatively correlated with LRVS in Utah, as shown in the middle row of Figure~\ref{fig:ps_lrvs_recom}.
This means that most of the time a larger positive partisan bias score incorrectly corresponds to a more Democratic-favoring plan (i.e., a plan with an LRVS below $0.5$). 

With the gubernatorial and combined data sets, the enacted plan has a positive partisan bias score, indicating that it favors the Republicans, but the LRVS of the enacted plan under those data sets is near the median LRVS score, that is, relatively neutral between the parties.
For the senate data the partisan bias is zero, which is supposed to indicate a fair plan, but the LRVS shows the plan heavily favors Republicans. 

For all three data sets, there aren't as many plans that have a negative (supposedly pro-Democratic) partisan bias. But all of those plans favor Republicans much more than the majority of the plans according to the LRVS score.

\subsubsection{Partisan Gini Score}
	 
The partisan Gini score, first introduced by \citet{Grofman}, is an unsigned score, meaning it does not indicate which party is supposed to be favored by a plan, just that the plan is considered unfair or asymmetric. \citet{DDDGMSS} show that the partisan Gini dominates other measures of partisan symmetry, including mean--median and partisan bias, meaning than when partisan Gini is zero (supposedly its ideal value), then the other measures of partisan symmetry also equal their supposedly ideal values. 

\citet{DDDGMSS} first observed with the 2016 senate election returns and a ReCom ensemble of 100,000 plans that every plan with a partisan Gini of zero also has all four districts with a Republican majority (corresponding to an LRVS greater than $0.5$). We see a similar result with the 2010 election returns and our larger ReCom ensemble. The gubernatorial and combined data sets have no plans with both a majority Democratic district (LRVS less than $0.5$) and a partisan Gini near zero (see the bottom middle and bottom right panels of Figure~\ref{fig:ps_lrvs_recom}). But with the senate data, there are a few plans with LRVS below $0.5$ and a partisan Gini near zero (bottom left panel of Figure~\ref{fig:ps_lrvs_recom}).

Because the partisan Gini is unsigned, it does not have a linear relationship with LRVS. Instead the relationship corresponds roughly to two line segments, one with a positive slope and one with a negative slope. The positive-sloping segment should correspond to Republican-favoring plans, with greater LRVS correlating to greater partisan Gini score (a pro-Republican gerrymander), and the negative sloping segment should correspond to Democratic-favoring plans, with lower LRVS correlating to greater partisan Gini score (indicating a pro-Democratic gerrymander). 

This expected double-line-segment structure is visible in all the data sets, but the line segments are fairly fat, indicating that for each partisan Gini value, there is a wide range of possible LRVS scores.  Specifically, a plan that was intentionally gerrymandered to have all four districts securely Republican (i.e., a high LRVS) could have the same partisan Gini score as a plan that gives a seat to the Democrats. This suggests that partisan Gini cannot necessarily distinguish a plan that was gerrymandered in favor of Republicans from a plan that was not.
For example, according to the senate data, the enacted plan is unusually favorable to Republicans, with an LRVS higher than $98.23\%$ of all plans in the ensemble, but it has a very low partisan Gini score.  

However, it is noteworthy that plans with an unusually low LRVS all have a higher partisan Gini, showing that Partisan Gini may still be able to identify a pro-Democratic gerrymander.

\subsection{Newer Measures}	 

\subsubsection{Average Absolute Partisan Dislocation}

The metric called \emph{partisan dislocation}, first defined by Eubank and Rodden and studied in detail by \citet{DeFordEubackRodden}, quantifies how much the partisan preferences in each voter's district differ from those in the voter's neighborhood. Larger values indicate that more voters are in districts that don't look like their neighborhoods. This score differs from the metrics discussed above because it is a voter-level (or precinct-level) metric. In order to construct a single metric which applies to the entire plan, the absolute value of the partisan dislocation for each precinct is averaged across all precincts. This is the \emph{average absolute partisan dislocation (AAPD)}.

\begin{figure}
\centering
\includegraphics[width = 12 cm ,height = 16 cm]{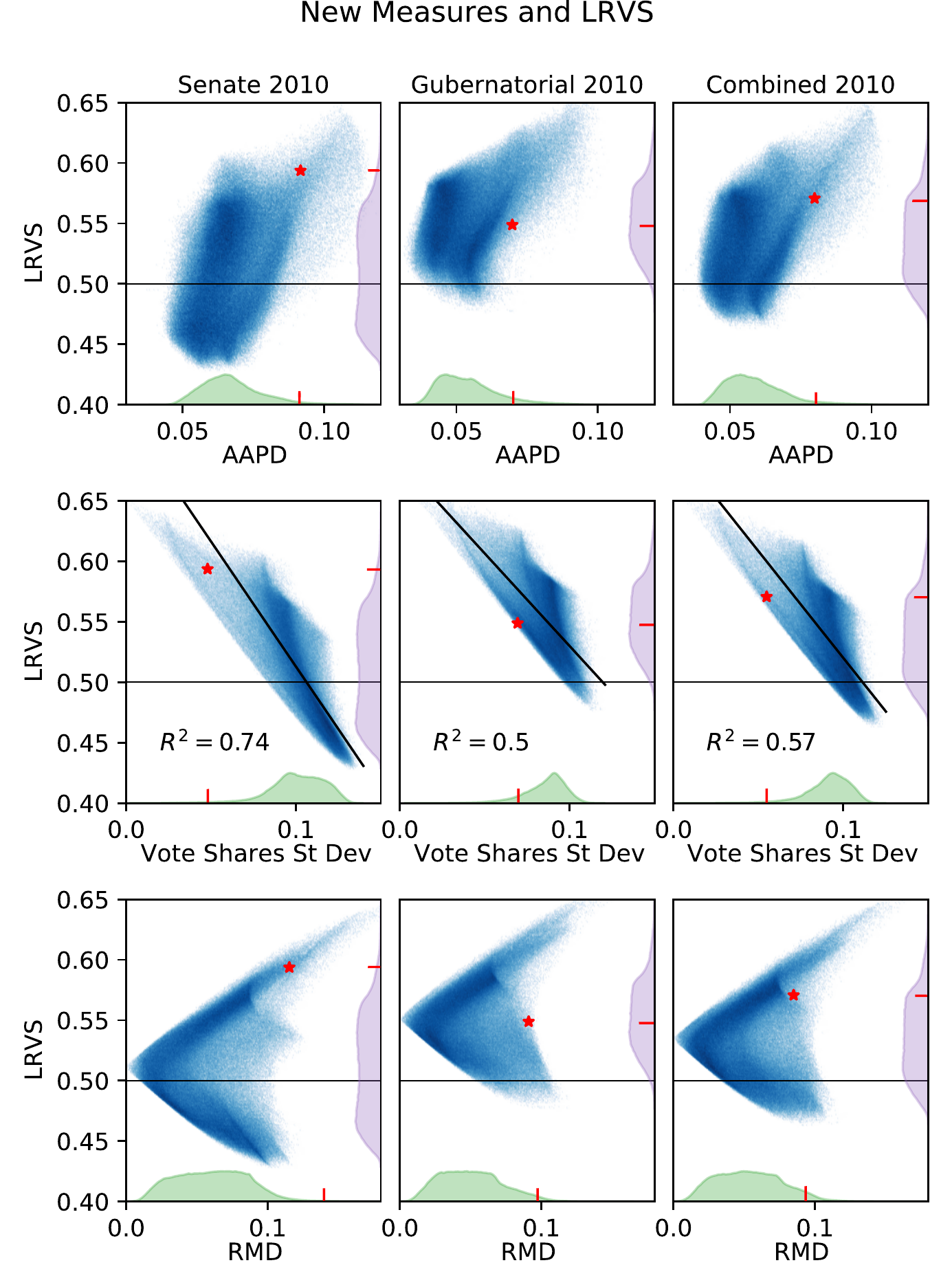}
\caption{Plots comparing the relationship between newer measures and LRVS in the ReCom ensemble. The marginal densities for the measure (green) and for LRVS (purple) are shown on the x and y axes respectively. 
}
\label{fig:other_lrvs_recom}
\end{figure}

Figure~\ref{fig:other_lrvs_recom} shows the relationship between average absolute partisan dislocation and LRVS in the ReCom ensemble. 
Since the average absolute partisan dislocation is unsigned, we cannot expect a good linear relationship between the partisan dislocation and the LRVS. Instead we should expect to see something like the shape of the partisan Gini plot for the senate data (see Figure~\ref{fig:ps_lrvs_recom}), where the upper and lower ends of the LRVS distribution would have greater partisan dislocation and the point of least partisan dislocation should occur somewhere in the middle. From the first row of \ref{fig:other_lrvs_recom}, this shape isn't prominent.
For most values of absolute average partisan dislocation there is a wide range of possible values of LRVS, indicating that most values of partisan dislocation give very little information about the impact of the plan on partisan advantage or LRVS, especially in the lower part of each plot.
In the top part we do see some slope, with a high LRVS  correlated with higher AAPD.  This suggests that AAPD might be able to detect a pro-Republican gerrymander, but probably not a pro-Democratic gerrymander. Moreover, there are some plans with a typical AAPD score and an atypical LRVS score and vice versa. This shows that the AAPD can be gamed by creating a plan with any desired LRVS and yet an AAPD that is typical of the ensemble.

\subsubsection{Standard Deviation of Vote Shares}
	
\citet{Magleby} used the standard deviation of vote shares in the four Utah congressional districts to conclude that the enacted plan in Utah is an ``egregious Republican gerrymander.'' His argument was that a small standard deviation is a hallmark of a cracking gerrymander, and the enacted plan had a much smaller standard deviation (using 2016 election data) than any of his 10,000 plans did.
	
The middle row of Figure~\ref{fig:other_lrvs_recom} shows scatter plots of the standard deviation with LRVS. All three data sets show a negative slope in the relationship between standard deviation and LRVS. Unlike the other measures we have examined, the negative slope here is expected if, as Magleby argued, a smaller standard deviation indicates a cracking gerrymander, because in Utah such a gerrymander could only hurt the Democrats. So a smaller standard deviation is expected to correspond to a more Republican-favoring plan (higher LRVS) and a greater standard deviation should correspond to a more Democratic-favoring plan (lower LRVS). 

For the gubernatorial data the standard deviation of the enacted plan is lower than all but $89.3\%$ of the plans in the ensemble, whereas the LVRS of the enacted plan is fairly typical of the ensemble ($54.9\%$ percentile).
But for the senate data, the standard deviation of the enacted plan is smaller than $98.7\%$ of the ensemble, agreeing with Magleby's analysis of 2016 data.

\subsubsection{Ranked-Marginal Deviation}

\citet{MattinglyNC} use an unsigned metric called \emph{ranked-marginal deviation (RMD)}, also called the \emph{gerrymandering index}, which is computed as the sum of the square of the differences between the sorted vote shares of a given plan and the medians of the sorted vote shares across the ensemble. This metric tracks fairly well with LRVS, as changing the vote share of Republicans in the least Republican district necessitates changing the vote share in other districts as well. Plans that are gerrymandered in either direction (as measured by LRVS) are associated with larger ranked-marginal deviation, which indicates that this metric is performing well and giving a fairly good signal.

The RMD is plotted with LRVS in the bottom row of Figure~\ref{fig:other_lrvs_recom}.
As expected with an unsigned measure, the scatter plots of RMD with LRVS all have the general shape of two line segments coming together in a point on the left side of the plot, with the height of the point near the median of the distribution for the LRVS. This suggests that small values of RMD correctly correspond to plans typical of the ensemble. Furthermore, the closer the RMD is to zero, the smaller the range of possible LRVS scores represented in the ensemble, suggesting that there is an LRVS score that is considered ``best'' according to the RMD metric, and to acheive LRVS scores far from this ``best'' score, the plan must also score poorly according to the RMD metric. The distinct, sharp point on the upper right means that plans with a very high percentage of Republican voters in the least Republican district (LRVS far from the median) are correctly flagged with a large value of RMD. The RMD of the enacted plan with the gubernatorial data is greater than $95.4\%$ of the plans in the ensemble, and for the 
senate data it is greater than $98.2\%$ of the plans in the ensemble.  

\subsubsection{Efficiency Gap and Buffered Declination}
	
The efficiency gap, described in \citet{McGhee}, tries to evaluate a plan by computing parties' ``wasted'' votes in each district. \citet{Veomett} showed that the efficiency gap suffers from a number of significant defects as a measure of gerrymandering. In the case of our ensembles in Utah, the efficiency gap takes only one value (approximately $0.19$ for the senate data) if the LRVS is above $0.5$, and it is very close to zero if the LRVS is less than $0.5$. Thus the efficiency gap gives no significant information beyond just counting the number of districts that have a Republican majority. 

\citet{Warrington} recently proposed another measure of gerrymandering called the \emph{declination}. Unfortunately declination does not make sense for plans with a single-party sweep, as often occurs in Utah. But there is a variant called the \emph{buffered declination} which does make sense even for plans with a single-party sweep.

In our ensemble buffered declination behaves like the efficiency gap for all plans that give all four seats to Republicans (LRVS above $0.5$).  Specifically, it gives  similar values for all such plans, regardless of whether the LRVS is barely over $0.5$ (more Democratic favoring) or is far above it (more Republican favoring).  
For plans that give one seat to the Democrats (LRVS below $0.5$), the buffered declination takes on a wider range of values, but it is negatively correlated to the LRVS for these plans.  In this it is very like mean--median and partisan bias, making a sign error by marking plans that favor Democrats as being more favorable to Republicans, and vice versa.

\subsubsection{Other Metrics}

A very common metric of gerrymandering is the total number of seats won by each party, but in Utah that metric carries no more information than whether the LRVS is above or below $0.5$.

One other metric used by \citet{MattinglyNC} is \emph{smoothed seat-count deviation}, or \emph{representativeness index}. The representativeness index compares how safe the most competitive Republican seat is compared to how safe the most competitive Democratic seat is. But in Utah many plans give all four congressional seats to the Republicans, which means the representativeness index is undefined. 

Finally, there are several other metrics discussed in the review by \citet{warrington_review} of gerrymandering metrics, but they are all just are minor variations of the metrics discussed above, so we do not treat them here.

\section{Discussion}

\subsection{Failure of Common Gerrymandering Metrics in Utah}

Because at least three congressional seats always go to the Republicans (with the 2011 precincts and 2010 election returns), only one seat can ever be contested.  The percentage of Republicans in the final, least-Republican, district, the LRVS, determines how difficult it is for the Republicans or Democrats to  win that last seat.  This  gives a very clear indication of how much a given plan favors Republicans or Democrats in Utah.   Our analysis shows that the classical metrics (partisan bias, mean--median, and partisan Gini) and many of the newer metrics (efficiency gap, partisan dislocation, and buffered declination) are uninformative at best, and misleading at worst, when applied in Utah.  

Of the four signed metrics, which are supposed to indicate which party is favored, partisan bias and the mean--median score make a sign error, incorrectly flagging plans with a higher LRVS as favoring Democrats and those with a lower LRVS as favoring Republicans.   Moreover, almost every plan that has a partisan bias or mean--median score of zero (supposed to be the ideal) strongly favors Republicans. A sign error for these two metrics in Utah was also observed by \citet{DDDGMSS} using the 2016 senate election returns.

The third signed metric, the efficiency gap, does not make a sign error, but it carries very little information, giving a single score ($0.19$ with the senate data) to all plans with four Republican seats and a score very close to zero for all plans that give one seat to the Democrats. 
The final signed metric, buffered declination,  also makes a sign error for half of the plans---those with LRVS below $0.5$. In that way it is like mean--median and partisan bias, giving misleading results.   For all plans with LRVS above $0.5$ it gives only a small range of values, essentially uncorrelated to any partisan advantage.  In that way it behaves more like the efficiency gap---not misleading, but also not very informative.

Most of the unsigned measures are expected to take their smallest values on plans that are perceived as optimal in some way (e.g., symmetric in the way they treat the parties or typical of acceptable plans), and take larger values for less desirable (asymmetric or atypical plans). These include ranked marginal deviation, partisan Gini, and average absolute partisan dislocation. 

Of these, the ranked marginal deviation seems to perform best. It has the good property that near zero (its lowest value, and supposedly the ideal) the LRVS is restricted to a range near its median, and for extreme values (much larger or smaller than typical) of the LRVS, the RMD is also large.  But it is not measuring exactly the same thing as the LRVS.  Specifically, larger values of the ranked marginal deviation do not guarantee that the LRVS is extreme.  The choice of which of these two metrics to use is then a question of policy---is it more important to consider how secure the last congressional seat is for the Republicans? or is it more important that each district have a Republican vote share that is typical of the ensemble?

The smallest values of the partisan Gini are not generally near the median LRVS, and are not even constrained to a small region (see, for example, the left panel of Figure~\ref{fig:ps_lrvs_recom}). Thus a small partisan Gini does not indicate a typical plan, as measured by the LRVS.

The lowest values of the average absolute partisan dislocation do slightly restrict the LRVS, but overall the dislocation seems to carry little information about how typical a plan is as measured by the LRVS.

Finally, the standard deviation of vote share is slightly different from the others in that it is not really designed as a test of partisan outcomes. Rather, it is a possible indicator of a cracking gerrymander, with smaller values indicating a more severe case of cracking. In Utah, a cracking gerrymander favors the Republicans, and the standard deviation does a fairly good job of tracking the LRVS with the expected negative correlation. 

\subsubsection{Summary: Evaluation of Metrics}

Aside from the LRVS itself, only the ranked marginal deviation (RMD) and standard deviation gave accurate information about Utah congressional plans. 
Low RMD corresponds to an LRVS that is typical of the ensemble,  but larger values of ranked marginal deviation do not necessarily indicate atypical plans. This is because it measures more than the LRVS---deviation of all district vote shares from the typical vote share for that district. 
The standard deviation can probably identify a cracking gerrymander and is correctly correlated with the LRVS.

\subsection{Unusually Good Scores for the Enacted Plan}

The enacted plan, when analyzed using senate 2010 data, performs exceptionally well on the common metrics in use at the time (partisan bias, mean--median, and partisan Gini). It has a mean--median score of $0.0021$ and a partisan Gini score of $0.0041,$ which are ``better'' (closer to zero) than $93.8\%$ and $97.6\%$ of the ensemble, respectively. The enacted plan also has a partisan bias score of exactly zero, but that is less remarkable, since the partisan bias score can only take on three values, and a large percentage of the plans in the ensemble had a partisan bias score of zero. 

Contrast these very good scores on the senate data and the metrics that existed in 2010 with the plan's poor scores on the metrics that were not in common use at the time, like standard deviation of vote shares, partisan dislocation, and ranked-marginal deviation. Similarly, the enacted plan scores poorly on the gubernatorial data set with all the measures, including the ones that were known at the time. This suggests that the enacted plan may have been chosen in part because it scored well on the 2010 senate data with the standard measures of partisan symmetry of the time.

\subsection{Was Utah Gerrymandered?}

Although the main focus of this paper is to evaluate how well the LRVS and common gerrymandering metrics perform  in Utah, we briefly explore the question of what our results suggest about whether the 2011 enacted redistricting plan in Utah was a partisan gerrymander. Our ability to answer that question depends on whether the ensemble we used is a good match to the redistricting principles that the enacted plan was intended to follow.  This is discussed in more detail in Section~\ref{sec:MCMC-methods}, but there is some reason to believe that the unconstrained ReCom method, which we used to generate our ensemble,  has a stationary distribution that is well matched to the redistricting principles adopted by the 2011 Legislative Redistricting Committee.  Moreover, the ensemble appears to be well mixed, suggesting that the ensemble is representative of that stationary distribution.

The answer to the gerrymandering question also depends on which data set best represents the distribution of voters at the time the plan was enacted. The two data sets, 2010 senate and 2010 gubernatorial, give different results.  

The 2010 gubernatorial data give the enacted plan an LRVS of $54.9\%$, which is close to the median of the gubernatorial LRVS scores in the ensemble. Indeed, it is more Republican than only $46.97\%$ of plans in the ReCom ensemble, which suggests that the enacted plan is typical of most of the plans in the ensemble with the gubernatorial data.
The standard deviation of the enacted plan with the gubernatorial data is lower than $89.3\%$ of the plans, which might suggest a cracking gerrymander, but it is not really extreme.  Finally, the RMD of the enacted plan for the gubernatorial data is greater than $95.4\%$ of the plans in the ensemble, making it something of an outlier.  The question for this data set then becomes whether an outlier RMD indicates a gerrymander when the LRVS is not an outlier and the standard deviation gives only a weak signal.  That is more of a question for the political scientists and the public: is it undesirable for a plan to deviate significantly from the typical distribution of vote shares over all the districts, even if the vote share in the lone competitive district (the LRVS) is typical?    

As discussed earlier in this paper, it seems likely that the 2010 senate returns give a more accurate reflection of the underlying partisan vote shares at that time than the special gubernatorial election.
The senate returns give the enacted plan an LRVS of $0.594$, which is a significant outlier in the ensemble. Indeed, it favors Republicans more than $98.23\%$ of the plans in the ensemble. The standard deviation for the enacted plan with the senate data is lower (more likely to be a cracking gerrymander) than $98.6\%$ of the plans in the ensemble. And the RMD for the senate data is also an outlier, being greater than $98.2\%$ of the plans in the ensemble.

In summary, if our unconstrained ReCom ensemble is well matched to the 2011 redistricting principles and if the the 2010 senate data set is an accurate reflection of the Utah electorate at the time the plan was enacted, then the three metrics that seem to be the most meaningful in Utah, namely LRVS, standard deviation, and RMD, all seem to suggest that the enacted plan is indeed  a partisan gerrymander.  The RMD does not indicate which party is favored, but suggests the result is extreme.  The LRVS indicates that the plan strongly favors Republicans. And the unusually small standard deviation suggests that this was a cracking gerrymander, cracking  the Democratic voters into multiple districts to reduce their ability to elect one Democratic Congressional Representative.  

\subsection{Application of LRVS to Other States}\label{other-states}

There are many other states which, like Utah, can only have at most one competitive congressional district. Any such state is amenable to an analysis with a score like the LRVS or the Least Democratic Vote Share. These states fall into a few categories. 

The first category includes states that, like Utah, are sufficiently small and sufficiently Republican that they could essentially never have two Democratic seats. For these states, which include Nebraska, Kansas, Oklahoma, Arkansas, and West Virginia, the least Republican district is the only possibly competitive district, so LRVS would be a useful indicator of gerrymandering. 

The second category includes states which are heavily Republican but have enough of a protected minority population to necessitate the creation of a majority-minority district to satisfy the Voting Rights Act. In these states, which include Mississippi, Louisiana, Alabama, South Carolina, the majority-minority district is Democratic, but there can be no more than one other district with a Democratic majority. Hence, the only competitive district is the second-least Republican district, so a score measuring the vote share in this district could be a useful indicator of gerrymandering. 

Finally, these methods could be used in states for which there could essentially never be more than one Republican district; these include Massachusetts, Connecticut, and Rhode Island. Because the least-Democratic district is the only  district that could elect a Republican in these states, the vote share in the least Democratic district would be a useful indicator of gerrymandering.

\section*{Acknowledgements}

We are grateful to Kristine Jones and Adam Shirey for their help with organizing, validating, and managing the data. Tyler Jarvis thanks Moon Duchin, Darryl DeFord, Greg Herschlag, Jonathan Mattingly, Jeanne Clelland, Greg Warrington, and Blake Esselstyn for helpful discussions. We thank Heidi Jarvis for thoughtful feedback and many useful suggestions for improvement.  We also thank the referees for insightful feedback and many helpful suggestions.

Funding for this project was paid for in part by a BYU Mentoring Environment Grant.

\section*{Data and Code Availability}
\noindent The data used in this paper are available at \url{https://github.com/tylerjarvis/UtahRedistricting}. The ensembles were generated using the \emph{Gerrychain} Python package, available at\\ \url{https://gerrychain.readthedocs.io/en/latest}.

\section{Disclosure Statement}

The authors report there are no competing interests to declare. 

\bibliographystyle{apalike}

\bibliography{gerrymander}

\appendix
\section*{Appendix: Finding a plan with two Democratic Districts}

To find a not-necessarily contiguous plan with two  Democratic districts out of four, we first found a plan that divides Utah into only two ``super'' districts of approximately equal population with one of the two having a majority of Democratic voters and then split that super district into two.  

One na\"ive approach for trying to find the super districts is to put the precincts with the greatest Democratic vote share into one super district and those with the least into the other super district.  This does not work because both resulting super districts have more than $50\%$ Republican vote share. But the population and the percentage of the population that vote in a precinct both vary widely across precincts, so we can increase the total Democratic vote share in the more-Democratic super district by swapping out a precinct with higher Democratic vote share and replacing it with a precinct of similar population that has somewhat lower Democratic vote share but more total voters.  Here is a toy example: if the super district has 10 democratic votes and 30 total votes, then swapping out a precinct with $100\%$ Democratic vote share but only 1 voter, and replacing it with a precinct having 4 Democratic voters out of 6 total voters gives a new vote share of $\frac{13}{35} = 37.1\%$, which is greater than the original vote share of $33.3\%$.

Repeatedly swapping such precincts in a greedy manner while complying with the the population constraint yielded a super district with a Democratic vote share of $50.2\%$, which we could then split into two regular districts, both with just over $50\%$ Democratic vote share.  The super district and the two split districts are far from being contiguous, with a broad scattering of precincts from across the state.  It seems very unlikely that such a split could be accomplished with contiguous districts.
\end{document}